\documentclass{emulateapj}
\slugcomment{{\sc Accepted to ApJ:} October 28, 2007}

\usepackage{natbib}
\usepackage{graphicx}
\usepackage{multirow}
\usepackage{amsmath}

\newcommand{\mean}[1]{\ensuremath{\left\langle #1 \right\rangle}}
\DeclareMathOperator{\Var}{Var}
\DeclareMathOperator{\Cov}{Cov}

\shorttitle{Hunting Galaxies to (and for) Extinction}
\shortauthors{Foster, Rom\'{a}n-Z\'{u}\~{n}iga, Goodman, Lada, Alves}
\begin{document}
\title{Hunting Galaxies to (and for) Extinction}

\author{Jonathan B. Foster, Carlos Rom\'{a}n-Z\'{u}\~{n}iga, Alyssa A. Goodman}
\affil{Harvard-Smithsonian Center for Astrophysics, 60 Garden Street, Cambridge, MA 02138}
\author{Elizabeth Lada}
\affil{Department of Astronomy,
216 Bryant Space Science Center,
P.O.Box 112055,
University of Florida,
Gainesville, FL 32611-2055}
\author{Jo\~{a}o Alves}
\affil{ Calar Alto Observatory
Centro Astron\'{o}mico Hispano Alem\'{a}n
C/ Jes\'{u}s Durb\'{a}n Rem\'{o}n, 2-2
04004 Almeria, Spain }

\begin{abstract}

In studies of star-forming regions, near-infrared excess (NIRX) sources---objects with intrinsic colors redder than normal stars---constitute both signal (young stars) and noise (e.g. background galaxies). We hunt down (identify) galaxies using near-infrared observations in the Perseus star-forming region by combining structural information, colors, and number density estimates. Galaxies at moderate redshifts (z = 0.1 - 0.5) have colors similar to young stellar objects (YSOs) at both near- and mid-infrared (e.g. Spitzer) wavelengths, which limits our ability to identify YSOs from colors alone. Structural information from high-quality near-infrared observations allows us to better separate YSOs from galaxies, rejecting 2/5 of the YSO candidates identified from Spitzer observations of our regions and potentially extending the YSO luminosity function below K of 15 magnitudes where galaxy contamination dominates. Once they are identified we use galaxies as valuable extra signal for making extinction maps of molecular clouds. Our new iterative procedure: the Galaxies Near Infrared Color Excess method Revisited (GNICER), uses the mean colors of galaxies as a function of magnitude to include them in extinction maps in an unbiased way. GNICER increases the number of background sources used to probe the structure of a cloud, decreasing the noise and increasing the resolution of extinction maps made far from the galactic plane.

\end{abstract}


\section{Introduction}

The near-infrared (NIR) has been a valuable window into star-forming regions, providing us with a variety of tools to study the process by which dark clouds coalesce into stars: luminosity functions in the $K$ band (2.2$\mu$m) can provide an accurate estimate of the initial mass function \citep[e.g.][]{Muench:2002,Stolte:2005};  studies of embedded clusters reveal important details of the processes by which stars form either in groups or in isolation \citep[e.g.][]{Lada:1991,Roman-Zuniga:2007}; and extinction mapping in the near-infrared allows precise determination of the column density structure of the cloud \citep[e.g.][]{Alves:2001,Cambresy:2002}.

Of particular use is the narrow range of intrinsic near-infrared colors of main-sequence stars (typically $H-K$ = 0 - 0.4 and $J-H$ = 0 - 1.0). This narrow range is due both to the near-infrared bands lying on the Wien tail of all stellar (hydrogen-fusing)-temperature blackbodies and to the relative paucity and uniformity of absorption feature\footnote{Principally H$^{-}$, but also CO features which are sensitive to surface gravity and produce a small split between dwarf and giant colors}. By assuming that all stars have the same intrinsic color we can do purely photometric extinction mapping. This is the heart of the Near Infrared Color Excess method (NICE) \citep{Lada:1994} and the Near Infrared Color Excess method Revisited (NICER) \citep{Lombardi:2001}.

Additionally, we can use the narrow range of stellar near-infrared colors to identify young stars by their near-infrared excess. The thermal contribution from their disk or envelope changes their color from that of a plain photosphere and places them in a certain region of a near-infrared color-color ($J-H$ versus $H-K$) diagram. This is the CTTS locus defined by \citet{Meyer:1997}. 

Unfortunately, a number of other astronomical objects also have intrinsic red colors, similar to CTTS. We refer to all intrinsically red objects as near-infrared excess (NIRX) sources. Of particular concern for studies of star forming regions are objects which are not either young-stellar objects (YSOs) or T-Tauri type young stars (TTS). These objects---brown dwarfs, a variety of evolved stars, galaxies, and AGN, are often studied in the near-IR in their own right, but in studies of star forming regions where we have limited information about their nature (perhaps only $J$,$H$, and $K$ photometry) they are contaminants which must be understood and identified in order to avoid common biases in these studies. For example, including intrinsically red objects in NICER produces an overestimate of the extinction.

In this paper we identify as galaxies a large number of NIRX sources at moderate redshifts (z = 0.1-0.5) in high-quality near-infrared images outside of the Perseus molecular cloud complex. After describing our observations and reduction in \S\ref{Obs} we present a detailed case that our NIRX sources are galaxies in \S\ref{ColorAndNumber}. In the heart of the paper we show how deep NIR images can help identify genuine YSOs in Spitzer observations (\S\ref{Clusters}) and how we can make use of these contaminating galaxies as additional probes for extinction mapping (\S\ref{GNICERSection}).

\section{Observations}
\label{Obs}

\subsection{Acquisition and Reduction}
\label{Data2}

The observations discussed herein are deep near-infrared observations of the Perseus Molecular Cloud complex, obtained as part of the COMPLETE (COordinated Molecular Probe Line Extinction and Thermal Emission) Survey \citep{Ridge:2006} and were introduced in \citet{Foster:2006}. These observations were obtained at the Calar Alto Observatory on the 3.5 meter, using the OMEGA 2000 wide-field NIR camera. We observed 6 fields in the cloud (four covering B5 and one on each of two dark clouds in the south-east: L1448 and L1451), as well as two control fields outside of the boundaries of the complex in regions of low extinction as estimated from a 2MASS-based extinction map \citep{Ridge:2006}. OMEGA 2000 offers a 0.45\arcsec pixel scale and low image distortion over the full 15.1\arcmin x15.1\arcmin \  field of view. Typical seeing conditions were $\sim$ 0.6-0.7\arcsec. We observed in three bands with short, dithered exposures. Total integration times per field were 45 minutes at $H$, 37 minutes at $K_s$ and 6 minutes at $J$. The control fields were observed with the same exposure times as the on-cloud frames, and reached roughly the same depths ($J$, $H$ $\sim$ 20 and $K_s$ $\sim$ 19)

Following standard instrumental flat-field and dark corrections, images were combined using the xdimsum package under IRAF. This package performed sky subtraction, cosmic-ray removal, and image co-addition. With our dither pattern, the end result is a trimmed image with roughly constant signal to noise over a 13\arcmin\ by 13\arcmin\ region. 

Photometry was performed on these trimmed images in two different modes. Source Extractor \citep{Bertin:1996} was run on the images with a 5$\sigma$ detection threshold and an estimated FWHM of 1.15\arcsec\ input (see \S\ref{Binaries}, Figure~\ref{Control04_FWHM}) to make a distinction between extended and point-like objects. Along with photometry in the three observed bands, a classification parameter, $C$, is returned by Source Extractor. $C$ is a number between 0 and 1, with 1 being a perfect point source. Our basic classification separates objects into point like or extended depending on their value of $C$ being larger or equal or smaller than 0.5.

\begin{figure*}
\includegraphics[scale=0.58]{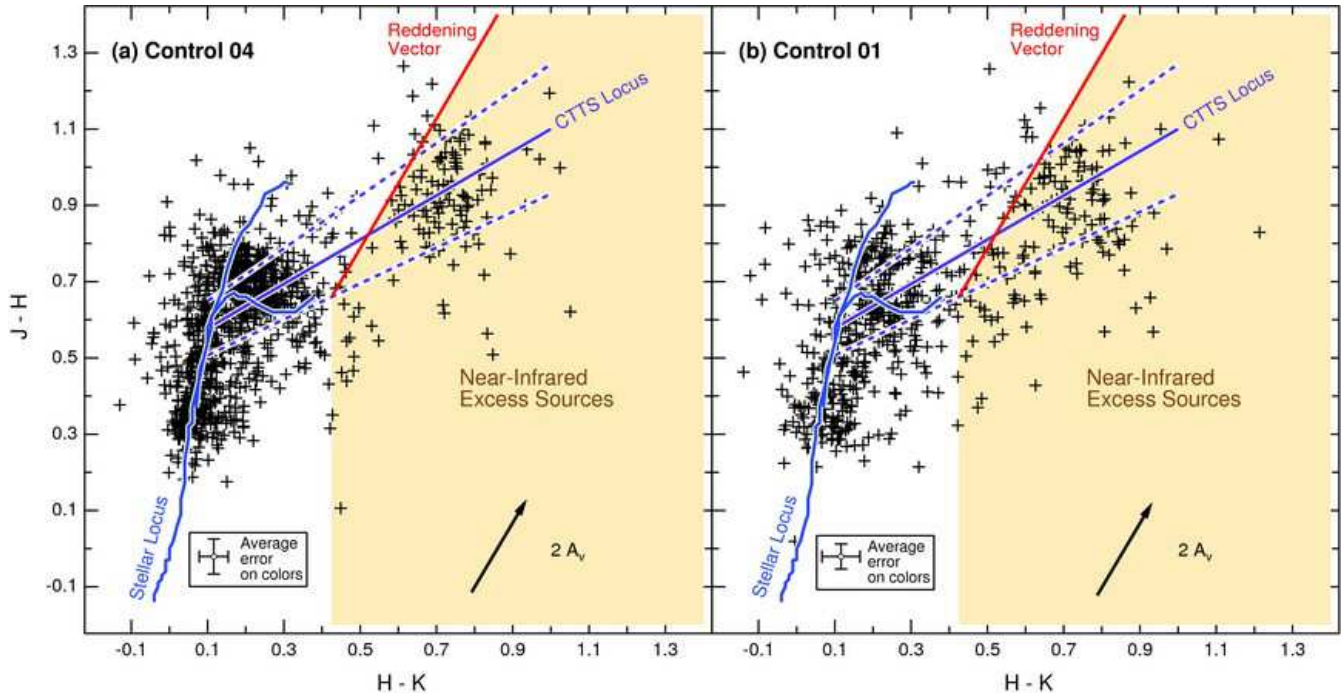}
\caption{Near-infrared color-color diagram for (a) Control 04 and (b) Control 01 in Perseus. The stellar locus is shown, consisting of an upper (giant) branch and lower (dwarf) branch, along with the classical T-Tauri locus from \citet{Meyer:1997}. Also displayed is a typical reddening vector  \citep{Rieke:1985}. We identify as Near-Infrared Excess (NIRX) sources all objects falling to the left of a reddening vector drawn through the tip of the dwarf stellar locus. Photometric errors will scatter some objects across this line in both directions.}
\label{ColorColorClump}
\end{figure*}

Source Extractor photometry produced a small number of very strangely colored (i.e. off the boundaries of the color-color diagram shown in Figure~\ref{ColorColorClump} either blueware or redward) objects which, on inspection, were objects at the edge of the frame with an unreliable sky estimate, stars in the wings of saturated bright stars, un-cleaned cosmic rays or bad pixels, or pieces of the diffraction spikes of bright stars misidentified as genuine objects. To eliminate such objects, we used the PhotVis package \citep{Gutermuth:2004} which performs DAOPHOT detection and aperture photometry with a fixed sky annulus and aperture size (here 5 pixels for a typical stellar FWHM of 2.5 pixels). The PhotVis package allows easy display and manual addition or removal of objects, so a hand-cleaned catalog was produced for each image. 

For the control fields, the catalogs from both programs were cross-referenced to find objects with detections in all three ($J$,$H$, and $K_s$) PhotVis catalogs and a detection in the $H$-band Source Extractor catalog. $H$-band was chosen for the classification image as it was the deepest and hence highest signal to noise image. For the data fields, the requirement of a detection in $J$-band was dropped, as differential extinction within the cloud caused this band to drop out for many sources. Magnitudes discussed herein are aperture photometry magnitudes from PhotVis, and are not necessarily accurate for extremely bright stars ($\ga$ 11 mag at $H$) or very extended objects ($\ga$ 2.5 $\arcsec$ at $H$). Our control fields contain only a few such objects, $\sim$ 10 total or 20 $\sq^{-1}$. Photometric errors are estimated from the sky noise in PhotVis. The control-field catalogs analyzed here were trimmed to include only sources with high enough quality colors to accurately gauge their position on the color-color diagram. This definition was $\sigma_{color} < $ 0.15 mag where

\begin{equation}
\sigma_{color} = \sqrt{\sigma_{H-K}^{2} + \sigma_{J-H}^{2}} = \sqrt{2\sigma_{H}^{2} + \sigma_{J}^{2}+\sigma_{K}^2}.
\end{equation}

Images were calibrated using 2MASS stars within the field. This produces a convenient way to establish a photometric system without the need for standard star observations and airmass corrections. The 2MASS point-source catalog \citep{Skrutskie:2006} was queried for all objects with high quality photometry (flag = AAA), cross-referenced with detected objects and used to determine a coordinate system. 2MASS stars with magnitudes between 11.5 and 13.5 were then linearly fit against detected magnitudes to establish a conversion between instrumental and 2MASS magnitudes.  The bright cut-off was chosen to eliminate potentially saturated stars or ones where the central pixel counts indicated the possibility of non-linear detector response. The formal uncertainty of this fit is not incorporated into the estimated errors for these stars, but constitutes an additional systematic error term. The $H$ and $K_s$ filters on OMEGA 2000 are the 2MASS filters. The $J$-band filter is somewhat broader at long wavelengths (a 50\% cutoff at 1.345 $\micron$ compared to 1.4 $\micron$ for the 2MASS filter), but is otherwise similar. Because of the similarly of filters and our calibration on 2MASS objects, we consider our observations to be on the 2MASS system and do not apply additional color-transformations.

\subsection {A clump of NIRX sources}

\begin{figure*}
\includegraphics*[scale=0.66]{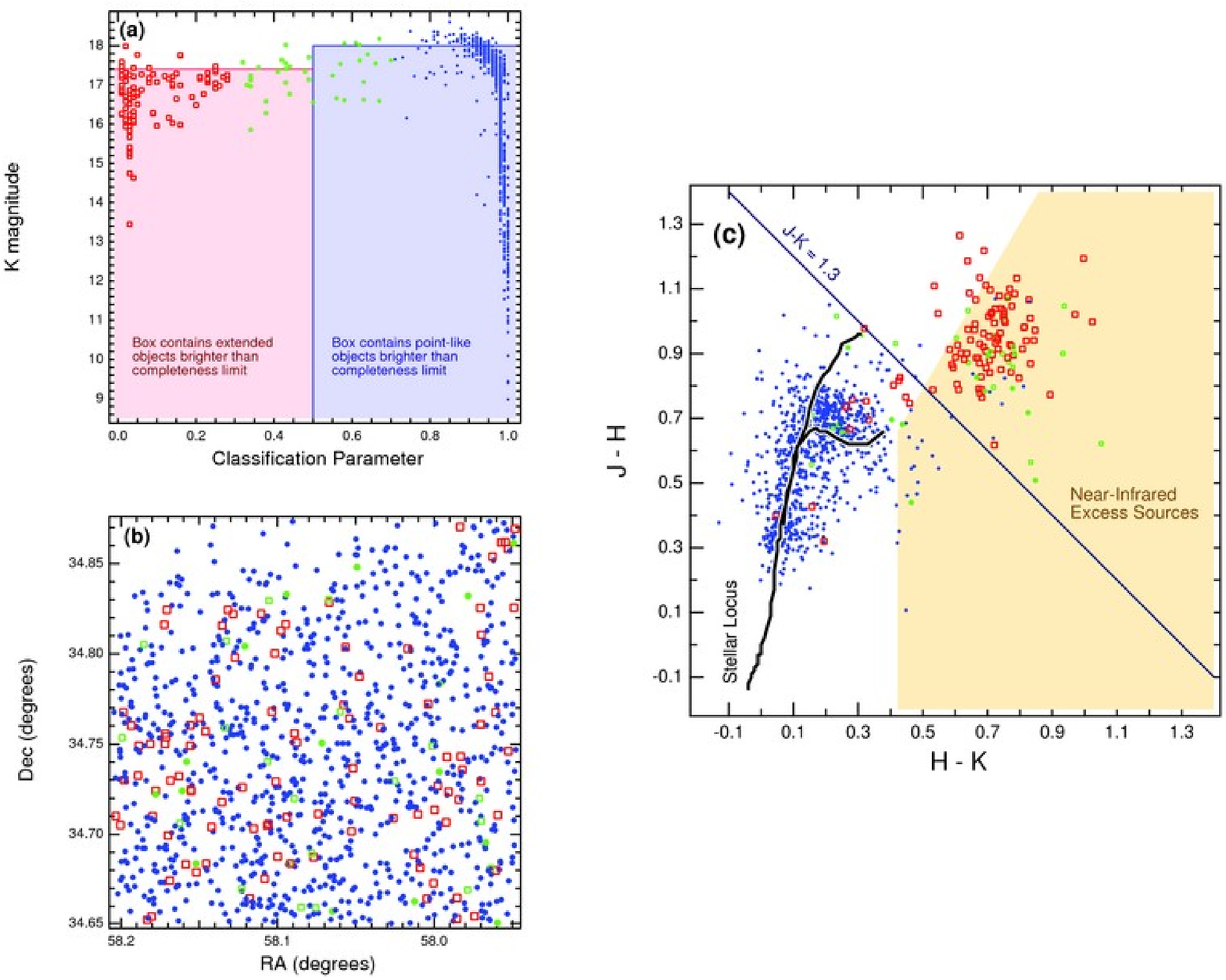}
\caption{Most near-infrared excess sources in Control 04 are extended objects. Panel (a) shows the classification of objects from Source Extractor; also shown are estimated completeness limits for extended (open/red) and point-like (closed/blue) object in $K$, at 17.4 and 18 magnitudes respectively. We highlight ambiguous classifications in green. Panel (b) shows that neither type of objects is clustered spatially. Panel(c) shows how this division captures the difference between stellar and NIRX sources in a color-color diagram. A typical galaxy/star line of $J-K$ = 1.3 is also shown, but would misclassify a number of sources.}
\label{Breakdown}
\end{figure*}

\begin{figure*}
\includegraphics*[scale=0.66]{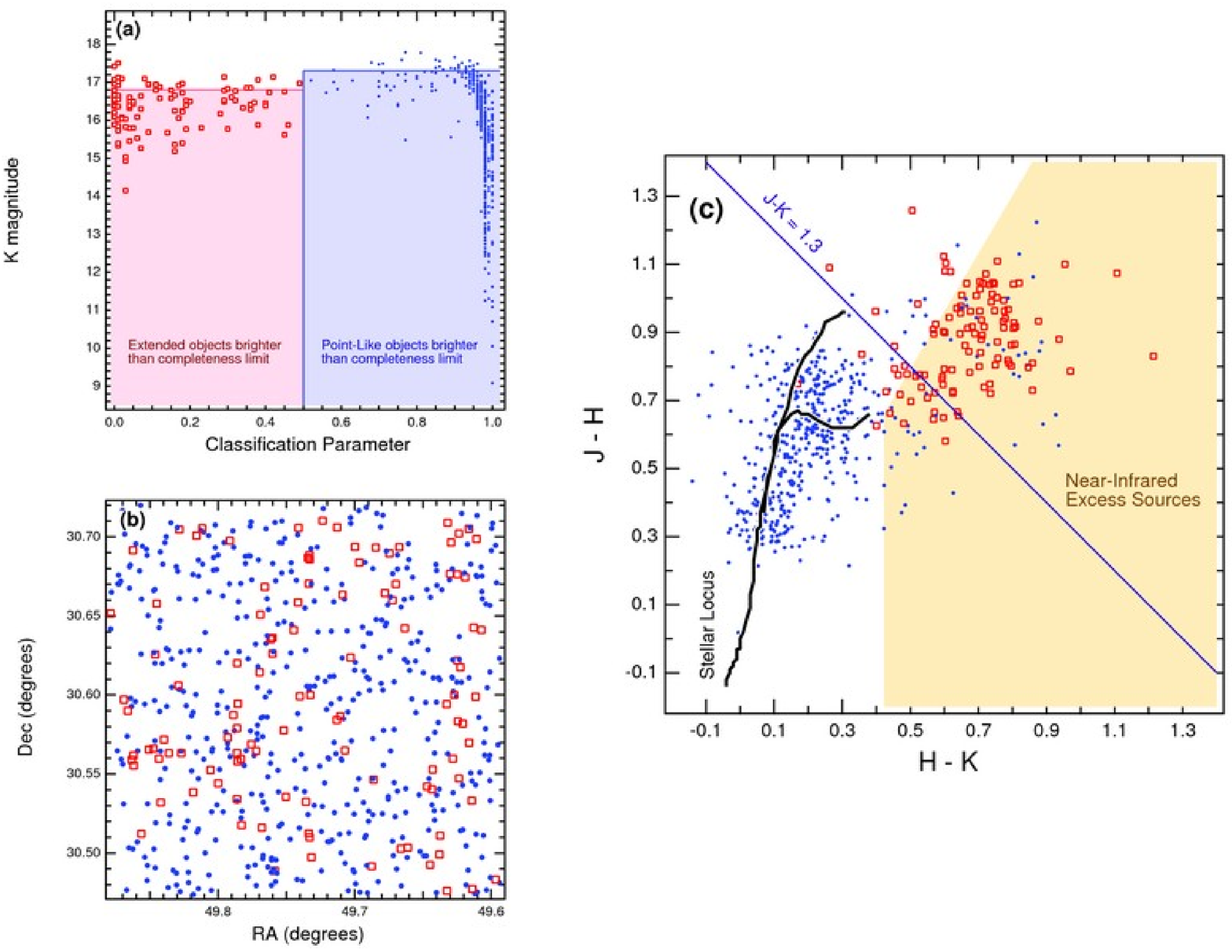}
\caption{The same breakdown as shown in Figure~\ref{Breakdown} for Control 01. The same behavior is seen as in Control 04, but the $K$ image is less deep, providing fewer sources and larger errors on the $H-K$ color, and thus decreasing the clustering of sources in color-color space.}
\label{Breakdown2}
\end{figure*}

The essential puzzle we address is illustrated in Figure \ref{ColorColorClump}; in both control fields there is a set of objects with fundamentally non-stellar colors -- colors which cannot be produced by any amount of reddening of a star. We define Near-Infrared Excess Sources (NIRX) in the following way: we offset from the tip of the main-sequence stellar locus (spectral type M6) by a buffer comparable to our photometric uncertainty and draw a reddening vector through this point. All objects to the right of this line are considered NIRX sources (see shaded region in Figure~\ref{ColorColorClump}). There is a distinct clump of objects in the same location in both of our control fields. This set of objects coincides roughly with the region of the diagram where young stellar objects (YSO) lie. Given that  Figure~\ref{ColorColorClump} shows \textit{control} fields far from the cloud that should not be littered with YSOs, what is the true nature of these objects? 

Source Extractor classifications are reliable for identifying bright galaxies with significant extent on the sky. This distinction becomes more difficult when there are image distortions toward the edge of a frame (a problem for many large detectors), in dense or confused fields, when galaxies are small or faint, or when there is significant nebulosity in the star-forming region. The last is likely to be always true for deep enough images due to the presence of ``Cloudshine'' \citep{Foster:2006}, that is, ambient galactic starlight reflecting off dust grains.

Faint sources constitute the bulk of our NIRX sources. Figures~\ref{Breakdown} \& \ref{Breakdown2} show sources from Control 04 and Control 01 in three different ways. In Figure~\ref{Breakdown}a, symbol types are assigned based on whether $C$ (classification parameter) is greater than or less than 0.5. Our estimated completeness limits (discussed in \S~\ref{CompEst}) are shown. The bulk of sources pile up at the extreme values of $C$. The faint objects in the middle (colored green) can not be assigned with the same level of confidence. Figure~\ref{Breakdown}b shows that there is no significant clustering of extended objects, nor are they more prevalent at the edges of the field, indicating that image distortions remain small towards the edge of the field. Figure~\ref{Breakdown}c shows that this division of extended versus point-like sources largely corresponds to a separation between sources near the stellar tracks and our clump of near infrared excess sources, though there are exceptions. To account for sources with marginal classification we consider the structural information from Source Extractor in combination with number density estimates of possible NIRX sources to securely identify the nature of our objects.

We analyze in detail just one (``Control 04'') of the control fields. This field is significantly deeper in $K$ due to a combination of rejected frames and poor weather in the other (``Control 01''). In Control 04 we have sufficient quality and depth to make a very secure case for the nature of the near infrared excess sources. 

\subsection{Completeness Estimate}
\label{CompEst}

To estimate the contribution of various objects with different luminosity distributions (e.g. YSOs or galaxies) to our survey requires knowledge of the completeness of our control field catalogs. This is complicated by the requirement that we detect a source in all three bands, that the error on its color be small ($<$ 0.15 mag), and by our source detection algorithm, which is a cross matching of the automated Source Extractor and aperture photometry from the Photvis package with some hand selection (\S\ref{Data2}).

\begin{figure}
\includegraphics[scale=0.55]{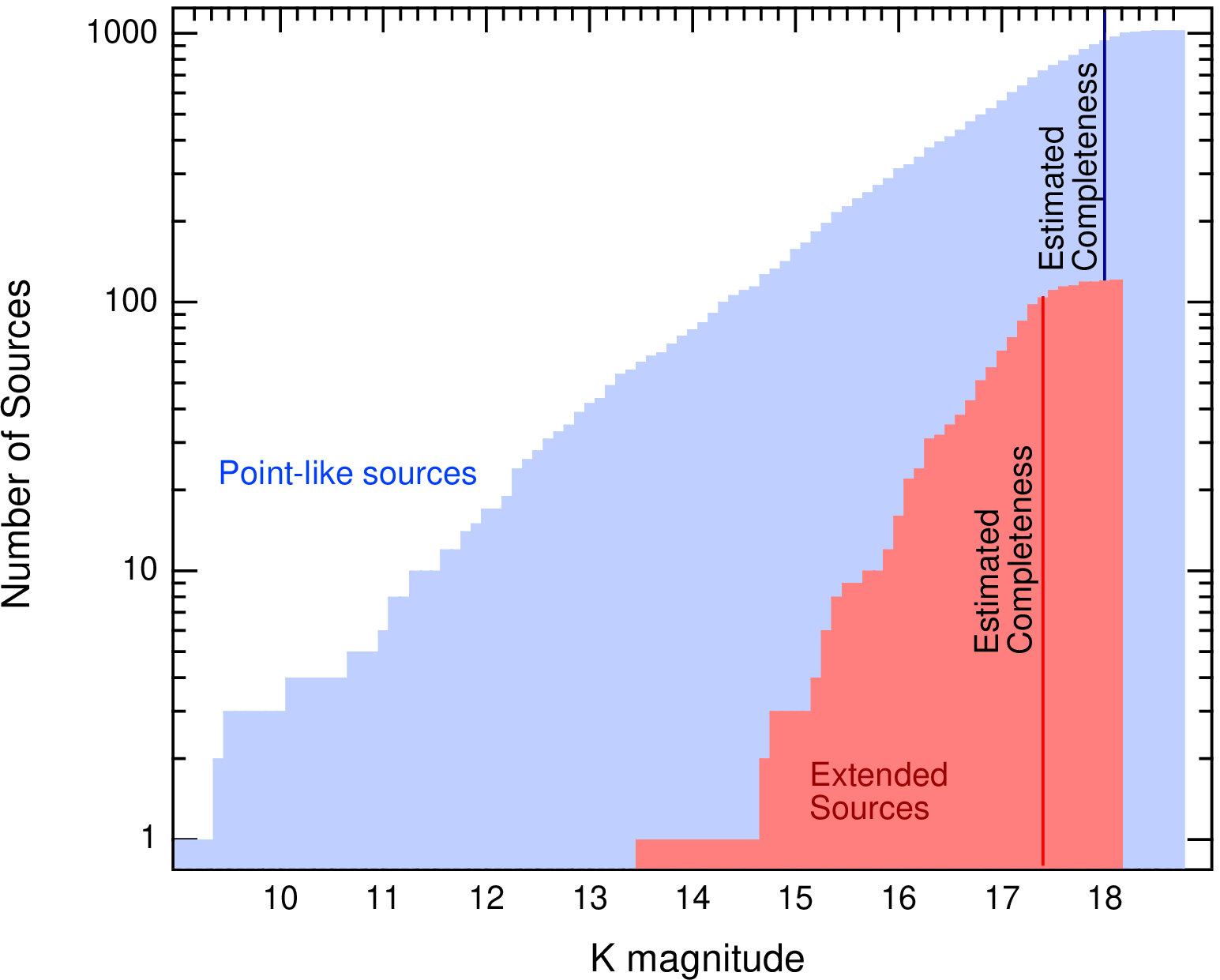}
\caption{Cumulative histogram and estimated completeness of point-like (blue) and extended (red) sources in Control 04.}
\label{Completeness}
\end{figure}

Rather than attempt to simulate this, we begin by examining the cumulative histogram for both extended and point-like sources in $K$. From the turnover in these plots (see Figure \ref{Completeness}) we consider our point-like sources to be complete to 18th magnitude and our extended sources to be complete to 17.4 magnitudes. Our estimated A$_V$ of 0.6 corresponds to an A$_K$ $\sim$ 0.07, and is small enough to be ignored when comparing number counts in our images to $K$-band magnitude distributions of candidate objects. We analyze the subset of sources brighter than these completeness limits in \S\ref{ColorAndNumber}

\section{Color description and Number Estimate of Candidate NIRX sources}
\label{ColorAndNumber}

To test the reliability of our structural information we seek a priori estimates for the number of NIRX sources we expect to see. Various point-like objects can have NIRX colors: young stars (already discussed), evolved stars (with a shell of cool ejected material), brown dwarfs (particularly young L-type dwarfs in the solar neighborhood) and quasars. Several kinds of sources can have extended profiles and be NIRX sources: normal galaxies at $z>0$ (normal local galaxies, composed only of stars and dust, should have either stellar or reddened-stellar colors), or exotic galaxies such as star-forming or AGN-dominated galaxies. Finally, visual binaries may or may not show up as extended depending on separation, and may or may not have stellar colors. 

\begin{figure*}
\includegraphics*[scale=0.57]{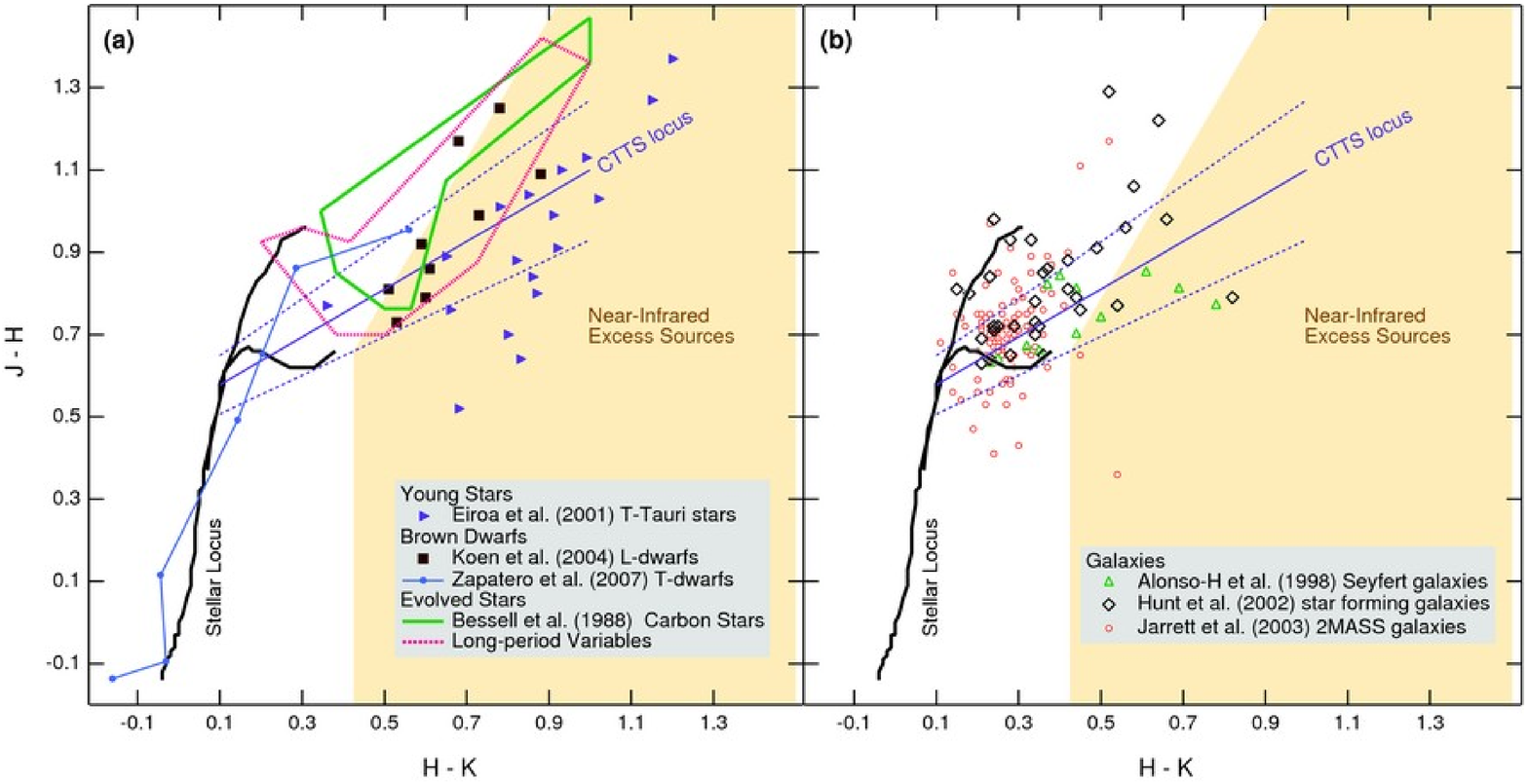}
\caption{The locations of candidate NIRX sources as taken from the literature. The thick dark line shows the stellar locus, while the purple-dashed line shows the classical T-Tauri locus. In (a) we show candidates within the galaxy: YSOs~(\S\ref{YSOs}), evolved stars~(\S\ref{AGB}), and brown dwarfs~(\S\ref{BDs}). In (b) we show extragalactic sources~(\S\ref{Galaxies} \& \S\ref{AGN})}
\label{Candidates}
\end{figure*}

In Figure \ref{Candidates} we show a $J-H$ vs $H-K$ color-color diagram with a literature compilation of colors of different galactic and extragalactic objects which are potential NIRX sources: the brightest (i.e. local) 2MASS galaxies from the compilation of \citet{Jarrett:2003}; galaxies with significant star formation and dust emission from the list of \citet{Hunt:2002}; L dwarf stars from the list of \citet{Koen:2004}; T-tauri stars with different ammounts of extinction and excess from the observations of \citet{Eiroa:2001}; colors of Seyfert galaxies from the list of \citet{Alonso-Herrero:1998}; an average sequence of T-dwarf colors constructed by \citet{Zapatero-Osorio:2007} from literature data; and the regions typically occupied by evolved stars: carbon stars and Long-Period Variable stars (with period $>$ 350 days) as defined by \citet{Bessell:1988}.

In the following section we estimate the contribution of the different kinds of objects listed above to our observed field. A summary can be found in Table~\ref{Estimates}.

\begin{deluxetable}{lll|l}
\tablecolumns{4}
\small
\tablewidth{0pt}
\tablecaption{Number Estimates for Objects in Control 04}
\tablehead{\multicolumn{3}{c}{Data} & \colhead {Estimated Contributions}} 
\startdata
Point-like & NIRX & 26 & 18 BDs, 2 Quasars, 0 YSOs, 0 AGBs \\
&	         normal & 818 & Normal Stars \\
\\
Extended & NIRX & 87 & 140 galaxies (not all NIRX) \\
&	  normal & 17  &  4 visual binaries \\
\enddata
\label{Estimates}
\end{deluxetable}

\subsection{Galactic candidates}

\subsubsection{YSOs}
\label{YSOs}
If an observed field is known to have recent or current star formation, NIRX sources are often taken to be YSOs. As explained by \citet{Lada:1992}, the combination of the stellar photosphere and disk place YSOs as near infrared excess sources, an example of which is a classic T-Tauri locus (CTTS locus) of \citet{Meyer:1997}. These sources are often used to trace the distribution of star formation within molecular clouds and to identify young embedded clusters \citep{Li:1997, Roman-Zuniga:2007,Gutermuth:2005,Ferreira:2007}. In our case, while there certainly are a number of young stars in the cloud regions, it seems unlikely that there is a large population of hitherto unknown young stars with disks present in both our control field which were selected to be outside the main regions of star formation in Perseus.  

Without a significant degree of clustering, the sources would not be a new young cluster, but a population of escaped sources from nearby star-forming sites, a situation that would raise current estimates of typical star formation efficiency or dispersal of young stars greatly. Assuming a distance of 250 pc to Perseus, Control 04 is $\sim$ 9 parsecs away from the nearest dense portion of Perseus (B5) and $\sim$ 14 pc away from the nearest site of clustered star-formation in IC348. If a YSO were born with a typical velocity of 1 km/s \citep{Kroupa:2003}, it would take between 9 and 14 Myr to reach the control field areas, which is larger than the typical period of NIR emission of a circumstellar disk ($\sim$ 5 Myr \citep{Haisch:2001}). Some small percentage of stars may leave their birth sites with significantly higher velocity, and so reach our control field before their disks evaporate. However, to constitute a significant number, the star formation rate in current clusters would have be dramatically higher in order that these velocity outliers would still be found in any number 14 pc away.

\subsubsection{Evolved Stars}
\label{AGB}
A star may also become a NIRX source at the end of its lifetime as it sheds its outer atmosphere, producing a shell of cold dust around the hot central star. This is the same basic mechanism by which  a YSO becomes a near infrared excess source. This is a broad category, encompassing planetary nebulae, carbon stars, AGB stars, various variable stars, and post-AGB stars. Recently, \citet{Lombardi:2006} suggested that the majority of contaminating near infrared excess sources in their study of the  Pipe Nebula are bright AGB stars, located at a similar distance near the galactic center (as evidenced by a narrow range in  magnitudes). With both our fields observed far from galactic center the contamination from these objects would be minimal. Additionally, our objects are much fainter than those reported by \citet{Lombardi:2006}. Other evolved stars are relatively rare, and are unlikely to contribute significantly.

\subsubsection{Brown dwarfs}
\label{BDs}
The near-IR colors of brown dwarfs are problematic. As illustrated by \citet{Stephens:2004}, the colors of brown dwarfs depend sensitively on the photometric system in which they are measured, since the same absorption features (H$_2$O) dictate the flux of the brown dwarfs, the design of NIR filters used from the ground, and the transparency of the atmosphere on a given observing night.

 \citet{Knapp:2004} show the colors of the L and T dwarfs in the MKO system, and \citet{Stephens:2004} provides the transformation between 2MASS and MKO. Within the uncertainties, it seems that L-dwarfs lie in the NIRX region of our color-color diagram, while T-dwarfs do not (see Figure \ref{Candidates}).

We estimate the expected number of L-dwarfs as follows. \citet{Chabrier:2000} estimate the galactic disk brown dwarf density at 0.1 pc$^{-3}$ and Figure~12 of \citet{Cruz:2003} provides a luminosity function for nearby ultracool objects (including M7 - L8) selected in a region of color-color space similar to the location of our NIRX sources. Even the brightest L-dwarfs are visible in our survey ($K$ = 18 mag) only if they are relatively nearby. For each magnitude bin we estimate the fraction of the total population in this bin, calculate the distance out to which this group is observable for our survey, multiply the derived volumes by the number density, and sum over all magnitude bins. This provides a very conservative estimate of the contamination, as we include many M7 - M9 stars whose colors lie partially overlapping the color locus of non-brown dwarf main-sequence stars. Our upper bound estimate is $\sim 18$ brown dwarfs, a significant contribution.

\subsubsection{Binaries}
\label{Binaries}
A large fraction of our stars are expected to belong to binary systems. An unresolved binary star may not lie on the unreddened stellar locus, but its color should still lie generally close to the locus as the average of two or more unreddened colors should stay close to the range of the separate components. Chance superpositions of two stars (i.e. a visual binary) could result in Source Extractor classifying such an object as extended. Such a superposition would be an extended object with a star-like color. 

\begin{figure}
\includegraphics*[scale=0.55]{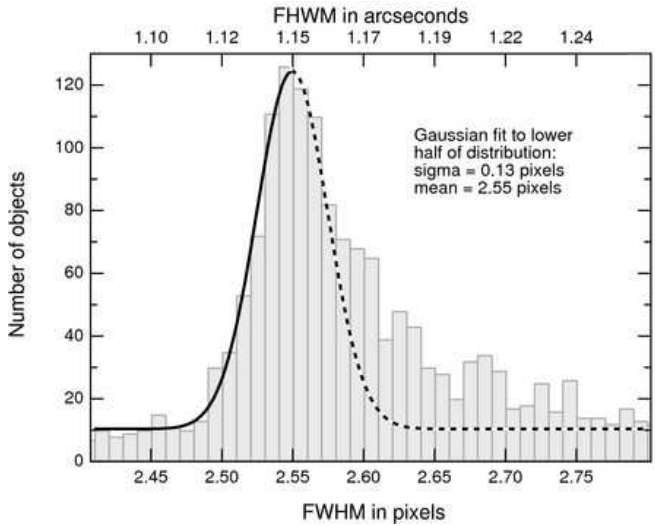}
\caption{Measurements of FWHM (by fitting a gaussian) of all objects in Control 04 $H$-band image down to 3 $\sigma$. The full range extends down to 0.1 and up to 20.2 and are almost certainly noise detections, diffraction spikes, or bad gaussian fits. The upper half of this distribution is contaminated by truly extended sources, so a gaussian is fit to only the lower half.}
\label{Control04_FWHM}
\end{figure}

In order to estimate the number of such superpositions we carried out a series of Monte Carlo Simulations.  To include the possibility that faint stars would distort the profiles of brighter objects, we lowered our detection threshold to 3$\sigma$ in $H$, producing ten times as many total sources as we analyzed in the rest of this paper. We measured the point-spread function (PSF) across the image by fitting each source as a 2-D gaussian. In Figure \ref{Control04_FWHM} we show the histogram of FWHM values. The high-end tail of the shown distribution is affected by truly extended objects, and a few odd noisy objects  produce a roughly constant tail of objects with a wide range of FWHM. We use this tail to determine a non-zero offset, above which we fit the width of the remaining distribution as $\sigma=0.13$ pixels.

\begin{figure}
\includegraphics*[scale=0.53]{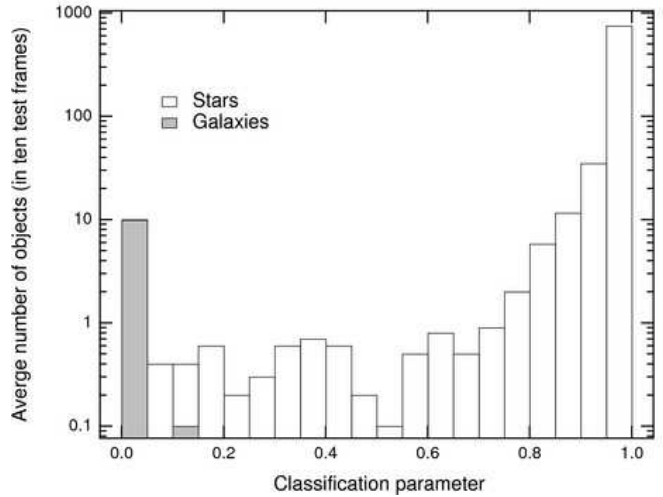}
\caption{The results of running Source Extractor on 10 test frames designed to reproduce the stellar density of Control 04. 10 input galaxies per frame are recovered, while a small number of other objects are classified as extended (Class $<$ 0.5). These objects are from overlapping point sources, and constitute roughly 0.4\% of the total objects.}
\label{SimulationHisto}
\end{figure}

We assumed that all these sources are point-like stars and randomly placed them back onto an empty image with uniform noise consistent with the noise in our real image. We input stars as gaussians with the distribution of FWHM values derived above---a mean of 2.55 pixels and a standard deviation of 0.13 pixels. To test recovery of truly extended objects, we also inserted a relatively bright elliptical galaxy sliced out of our data frame ten times into each image. To simulate the additional selection effects required to produce our catalog of sources we selected the 1024 objects with lowest photometric uncertainty (the same number of objects as in our real catalog) and retrieved their classification parameter value. The average number of objects assigned each classification parameter is shown in Figure \ref{SimulationHisto}.

We generated ten simulated fields and ran Source Extractor with the same detection parameters used in the original image. In all cases pasted elliptical objects are recovered and properly classified as extended. However, we also found some other sources which are classified as extended. When we visually inspected these objects in the simulated fields, we confirmed that they were indeed superpositions of two points sources.  The average number of such objects in all simulations, is 4$\pm$2 or 0.4\%.

\subsection{Extragalactic Candidates}

\subsubsection{Quasars and AGN}
\label{AGN}
Active galactic nuclei could be classified as point-like or extended depending on the brightness of the host galaxy. \citet{Hewett:2006} have produced color predictions for the WFCAM photometric system and conversions between these and the 2MASS colors. From these colors it appears as if quasars with z $<$ 6 might show up as near infrared excess sources.

A careful calculation of the detection of quasars in the near infrared has been performed by \citet{Maddox:2006}. We use their result for a $K$-band survey to 18.5, and assuming their most restrictive model for the combined quasar and galactic light (equivalent to being a point-like source), the estimate from their Table B5 is roughly 35 quasars per square degree up to 18th magnitude, or roughly 2 in our 13 arc minute by 13 arc minute control field.

\subsubsection{Galaxies}
\label{Galaxies}
Surveys of galaxies in the near infrared are typically done in the $K$ and $J$ bands. The reason is that $J-K$ has long been used as a simple galaxy-star discriminator, using a criterion where any object with $J-K >>1$ is a galaxy (e.g. \citet{Saracco:1999}). This distinction corresponds to a diagonal line in our color-color diagrams (see Figures~\ref{Breakdown}c~\&~\ref{Breakdown2}c), and agrees marginally well with our structural classification. The 2MASS survey contains $J$, $H$ and $K$ data, but galaxies in that survey are limited to objects much brighter than in our field. Few other surveys use $H$ band, although this is beginning to change (c.f. UKIDSS \citep{Lawrence:2006}, \citet{Metcalfe:2006}). Therefore, we consider number density estimates, a clustering analysis, and theoretical k-corrections to understand the properties of our extended objects.

Our density of extended objects agrees moderately well with deep extragalactic number density estimates. \citet{Elston:2006} provide data from the FLAMINGOS Extragalactic Survey (FLAMEX) and compare number densities with previous surveys. Integrating over their differential number counts down to magnitude 17.3 (our completeness of 17.4 with A$_K$ = 0.07) gives roughly 3000 galaxies per square degree, or roughly 140 galaxies detected in our field, compared to the 104 extended objects we identify. A second estimate is available from the GALAXYCOUNTS\footnote{http://www.aao.gov.au/astro/GalaxyCount/} tool available from \citet{Ellis:2007} which produces estimated number counts and variances in several bands including K. This tool predicts 150 $\pm$ 27 galaxies for our survey's characteristics. This result is almost 2$\sigma$ discrepant with our numbers, but is quite sensitive to our completeness estimate and small variations in magnitude definitions. If we were only complete to a depth of 17.1 then the estimate from \citet{Ellis:2007} of 121 $\pm$ 24 galaxies is 1$\sigma$ consistent with our number.

\begin{figure*}
\includegraphics*[scale=0.50]{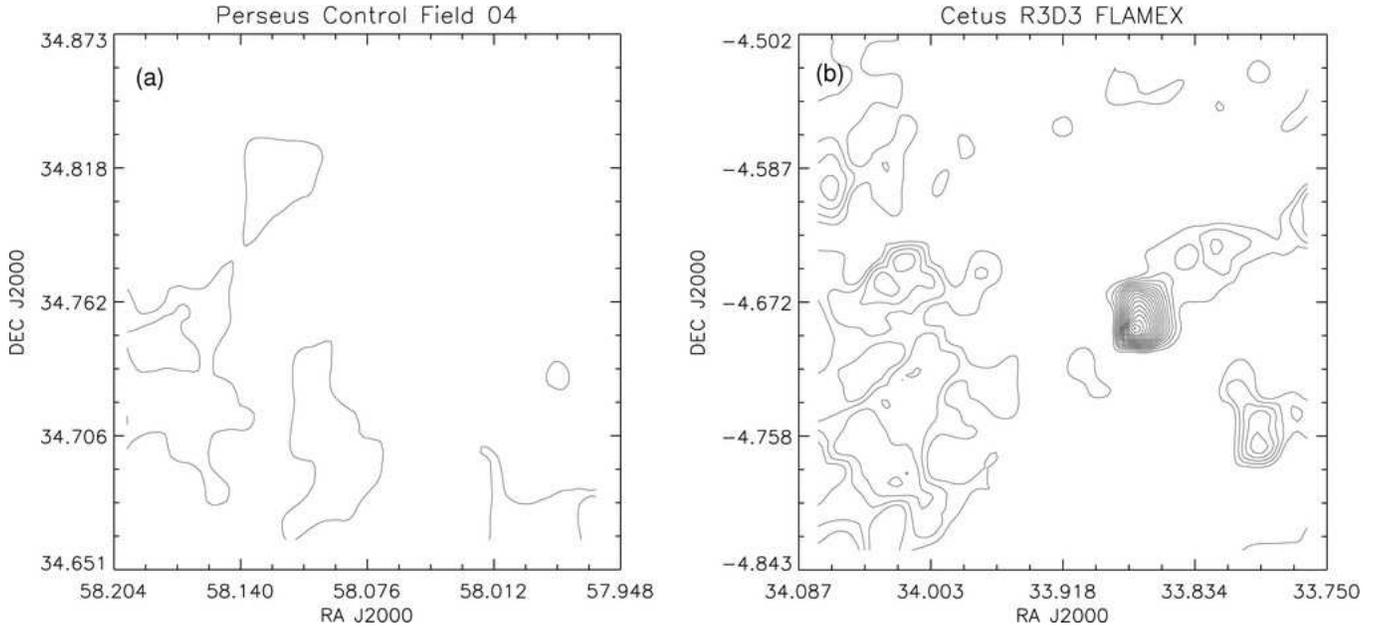}
\caption{The degree of clustering, as measured by 6th nearest neighbor distances translated into local surface densities (see text), of extended NIRX sources (in our field Control 04) and for a known galaxy cluster in the FLAMEX survey \citep{Elston:2006}. Contours are at 0.5 objects per square arcminute. The low degree of clustering in our field indicates we are not observing a galaxy cluster.}
\label{Concentration}
\end{figure*}

In order to test that our control field was not pointing by coincidence at a cluster of galaxies, we compared the surface density of extended objects ($C <0.5$ and $K<17.4$) with that of galaxies in one of the FLAMEX survey fields \citep{Elston:2006}. We selected from their Cetus catalog one region named R3D3 with central coordinates ($\alpha$(2000),$\delta$(2000))=(33.91840,-4.67243) which is known to have a prominent galaxy cluster; we selected galaxies with $K_s<17.4$ and ellipticities ($1-b/a$) larger than 0.2, in order to avoid including any significant numbers of stellar profile objects, which resulted in a total of 305 galaxies. To compare the surface densities, we calculated 6th nearest neighbor distances, $d_6$, for objects in both lists, which can be translated to local surface densities as $5/(\pi d_6^2)$ \citep[c.f.][]{Casertano:1985,Ferreira:2007}. In the Cetus-R3D3 and the Control 04 fields the mean densities of objects are $1.06\pm 1.30$ and $0.53\pm 0.25$ objects per square arcmin, while the maximum densities are 11.14 and 1.33 objects per square arcmin, respectively. In Figure~\ref{Concentration} we show maps of 6th neighbor surface densities for these fields in steps of 0.5 objects per square arcmin. While the nearest neighbor map clearly identifies the Cetus R3D3 cluster, the map for Control 04 shows no evidence of a cluster.

Finally, we seek to understand why our (non-cluster) galaxies form a relatively tight clump in the near-infrared color-color diagram. Down to $K$=17.4 mag, the depth of our observations, a significant number of galaxies would be expected to  be at moderately high redshift. Two effects can contribute to change the colors of distant galaxies. The first is the $k$-correction, which means that at higher redshift our near-infrared bands are observing intrinsically bluer portions of the galactic spectrum. The second is intrinsic galactic evolution --- different star formation rates, for instance. 

Due to the paucity of $H$-band deep extragalactic imaging, we do not have a way to compare colors of genuine red-shifted galaxies with those of the objects in our NIRX clump. Thus, we rely on theoretical $k$-corrections to understand the colors of our extended objects. We consider two different k-correction models, that of \citet{Mannucci:2001} based on near-infrared template spectra, and the HYPERZ templates by \citet{Bolzonella:2000}, which are visual spectra extended in wavelength using a population synthesis code. Both these models are compared in \citet{Hewett:2006}, who also carefully consider their transformations between different filter systems,  and we use their transformations into the 2MASS photometric system\footnote{Our J-band magnitudes are taken with a filter which is significantly different from the 2MASS one. Therefore, although we calibrate to the 2MASS system, this transformation is inexact for non-stellar objects}. Also, we apply a constant extinction of A$_K$=0.07 to both models to account for the extinction present in our control field.

\begin{figure*}
\includegraphics[scale=0.58]{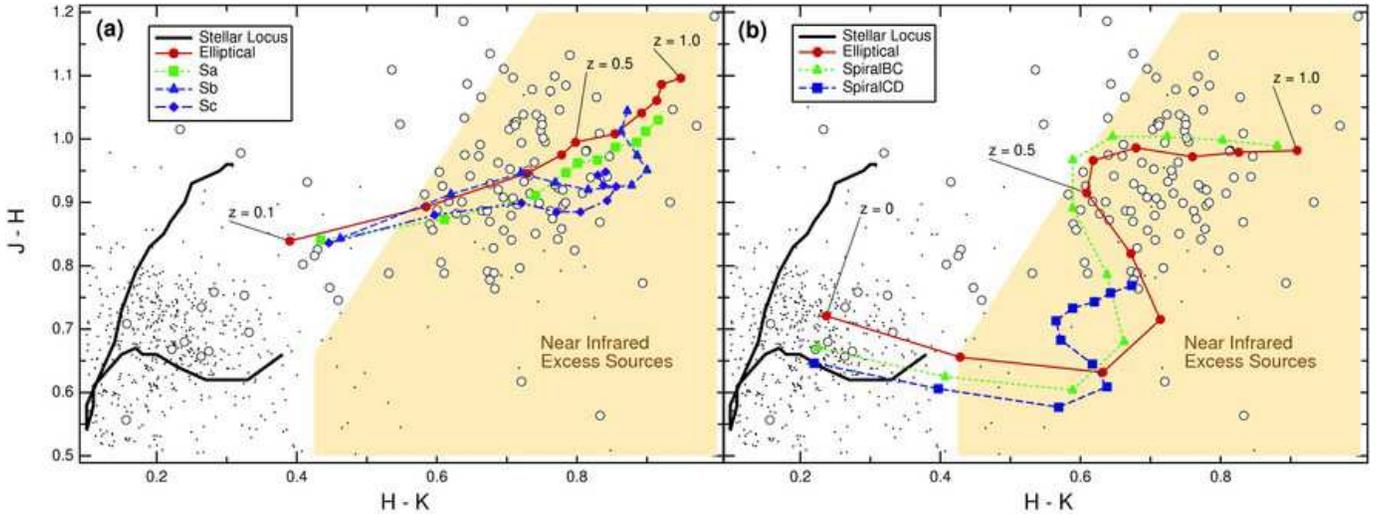}
\caption{K-correction models for different galaxy types overlain on control field color-color diagram. Again, open symbols are extended and points are point-like objects. Each point on the (colored) curves corresponds to a redshift increase of z = 0.1. (a) shows colors derived from Kinney-Mannucci spectra transformed into the 2MASS system from \citet{Hewett:2006} and (b) shows colors derived from HYPERZ templates transformed into the 2MASS system from \citet{Hewett:2006}. Note that the curves in (a) begin at z = 0.1. Both $k$-correction models have been shifted up along the reddening vector by 0.7 A$_K$ the estimated extinction in our control field.}
\label{Redshifts}
\end{figure*}

The comparison between our data and the $k$-correction models is shown in Figure~\ref{Redshifts}. We found that there are fairly significant differences in the shapes of the two models, but most of our extended objects can be explained as galaxies at moderate redshift. To quantify this, we bin our extended objects by $K$-band magnitude into three magnitude bins from 14 to 17. For each bin, we measure the median and 1$\sigma$ dispersion of $H - K$ and $J - H$. From \citet{Songaila:1994} we obtain the expected median redshift for each $K$-band magnitude bin. This redshift is converted into an $H-K$ and $J-H$ estimate using both the Kinney-Mannuci models and the HYPERZ models and the results shown in Table~\ref{ColorComparison}. The HYPERZ templates do not match our observed colors (missing most severely in the suspect $J$ band), but the Kinney-Mannucci templates appear to explain well the colors of our NIRX sources as being due to a distribution of normal galaxies of various types at redshifts between 0.1 and 0.5.

We are not able to distinguish this population of normal galaxies from more exotic types of AGN-dominated or starburst galaxies based on NIR colors alone. However, since our number density estimate for normal galaxies is larger than or consistent with our observed number density, and we understand the color-color clumping behavior as a consequence of the $k$-correction, we conclude that we are not seeing a significant population of exotic galaxies.

\begin{deluxetable*}{cclcc|ccc|cccc}
\tablecolumns{12}
\small
\tablewidth{0pc}
\tablecaption{Comparison of galaxy colors with models.}
\tablehead{
\multicolumn{5}{c}{Data properties} & \multicolumn{3}{c}{HYPERZ model} & \multicolumn{4}{c}{Kinney-Mannuci model} \\
\colhead{Mags} & {Median(z)} & & \colhead{Median}
& \colhead{$\sigma$} & \colhead{Elliptical} & \colhead{Sbc} & \colhead{Scd} &
\colhead{Ellip.} & \colhead{Sa} & \colhead{Sb} & \colhead{Sc} }
\startdata
14-15 & 0.18
& H - K & 0.53 & 0.10 & 0.627 & 0.588 & 0.570 & 0.543 & 0.570 & 0.583 & 0.562 \\
& & J - H & 0.82 & 0.10 & 0.701 &  0.672 & 0.647 & 0.881 & 0.865 & 0.897 & 0.87 \\
15-16 & 0.28
& H - K & 0.70 & 0.14 & 0.739	 & 0.688 & 0.665 & 0.706 & 0.720 & 0.704 & 0.701 \\
& & J - H & 0.90 & 0.09 & 0.766 & 0.732 & 0.668 & 0.938 & 0.905 & 0.941 & 0.896 \\
16-17 & 0.35
& H - K & 0.73 & 0.16 & 0.729 & 0.688 & 0.665 & 0.756 & 0.765 & 0.747 & 0.748 \\
& & J - H & 0.95 & 0.14 & 0.835 & 0.801 & 0.692 & 0.962 & 0.931 & 0.938 & 0.891 \\
\enddata
\label{ColorComparison}
\tablecomments{ On the right are observed median colors within set magnitude bins. By using the measured magnitude-redshift relation of  \citet{Songaila:1994} we predict colors using the two different sets of models introduced in the text for various galaxy types.}
\end{deluxetable*}

\section{Implications for Cluster Studies}
\label{Clusters}
\subsection{Cluster Studies in the Near-infrared}

Our analysis has a direct implication in the study of star forming regions. We consider data from the Rosette Molecular Cloud \citep{Roman-Zuniga:2007} which is of insufficient quality (due to instrumental problems) to make a good distinction between extended and point-like objects. However, it illustrates how contamination by galaxies could be important in cluster studies.

$J-H$ vs. $H-K$ color-color diagrams are a common tool used to determine which objects in a certain region have circumstellar emission. In the classical picture, the data are plotted in the color-color diagram and every object falling to the right of the reddening band of the zero age main sequence (e.g. $J-H < 1.692(H-K) + B$, with $B$ accounting for a typical color error buffer) is considered a candidate young star. However, our study shows that a significant fraction of these objects, especially faint ones, may be galaxies, rather than YSOs.

In many cases the quality of the observations can add to the problem. In order to make a good separation of elongated or fuzzy objects which might be extragalactic contaminants in a field populated with young stars using automated software, seeing and resolution need to be good. Still, in some cases, even with good quality observations, some extragalactic objects might not be detectable if their elongations are small, and thus some sources may pass as genuine young stars. The best solution, in most cases, is to have a good quality control field near the region of interest and perform in it a similar analysis to the one we performed in the Perseus field, which can give an estimate of the level of contamination in the field. Other factors like extinction might also have to be taken into account especially if the star forming region is very young.

\begin{figure}
\includegraphics[scale=0.53]{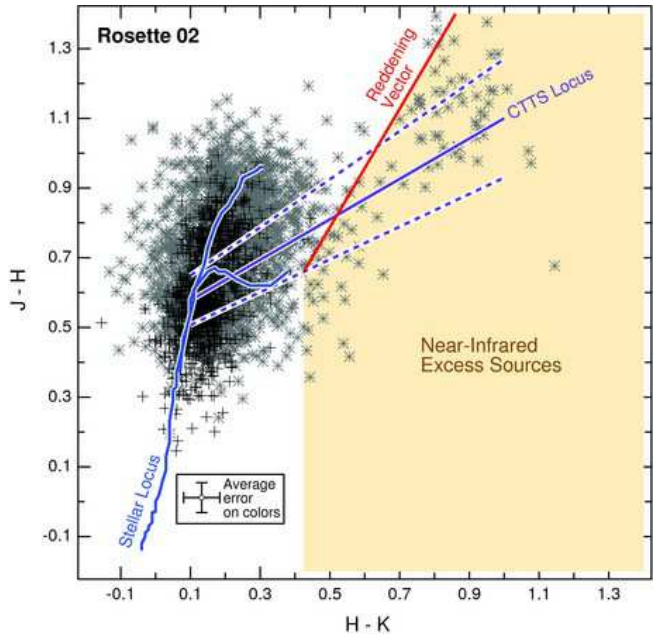}
\caption{Near-infrared color-color diagram for a control field in the Rosette nebula from \citet{Roman-Zuniga:2007}, with the same lines and regions as Figure~\ref{ColorColorClump}. Unable to use structural information to eliminate the clump of NIRX sources, \citet{Roman-Zuniga:2007} were forced to trim at K = 15.75 magnitudes (points fainter than this limit are shown as grey stars), roughly 2 magnitudes brighter than their detection limit. This tightens the dispersion on stellar colors and eliminates most of the NIRX sources.}
\label{ColorColorClump_Rosette}
\end{figure}

In their study of the Rosette Molecular Cloud, \citet{Roman-Zuniga:2007} decided that for sources fainter than $K=15.75$ mag they could not distinguish genuine NIRX sources from galactic or extragalactic contaminants, or from distorted objects produced by instrumental problems. Shown in Figure~\ref{ColorColorClump_Rosette} is one of their control fields displayed as in Figure~\ref{ColorColorClump} showing both faint (grey) and bright (black) sources. A similar set of NIRX sources is seen, mostly in faint objects. However, using a slightly larger area for possible NIRX sources than we use in this paper, they found that even after cutting out faint objects their two control fields (carefully selected to be located outside of the Rosette Cloud) still contained some NIRX sources  They used this number of NIRX sources to define a background level, which helped them to determine which regions of the Rosette have young cluster formation (regions with surface densities of NIRX sources above the defined background). If they had not corrected by this background population, then larger area of their cloud fields would present a high density of infrared excess sources and might have passed as star forming regions.

\subsection{Identifying YSOs with Spitzer data and High-quality NIR Data}

Galaxies masquerading as YSOs have been a significant headache for the analysis of Spitzer observations of star-forming regions, and contamination by galaxies may set the limit for identifying the lowest luminosity YSOs from these data alone \citep{Harvey:2007}. Near-infrared imaging is but one source of additional data which may be brought to bear on this problem. We combine our observations with data from the Cores to Disks (c2d) Legacy program survey of Perseus presented in \citet{Jorgensen:2006} to check our classification scheme and to demonstrate the value of adding high-quality near-infrared imaging to these rich datasets. As we will show, additional constraints come mainly from structural, rather than color information.  Unfortunately, the refined distinction between point-like and extended objects provided by NIR data is still not a definitive way to separate YSOs from galaxies, as YSOs may exhibit outflows or nebulosity, and distant galaxies may remain unresolved.

\begin{figure*}
\includegraphics*[scale=0.60]{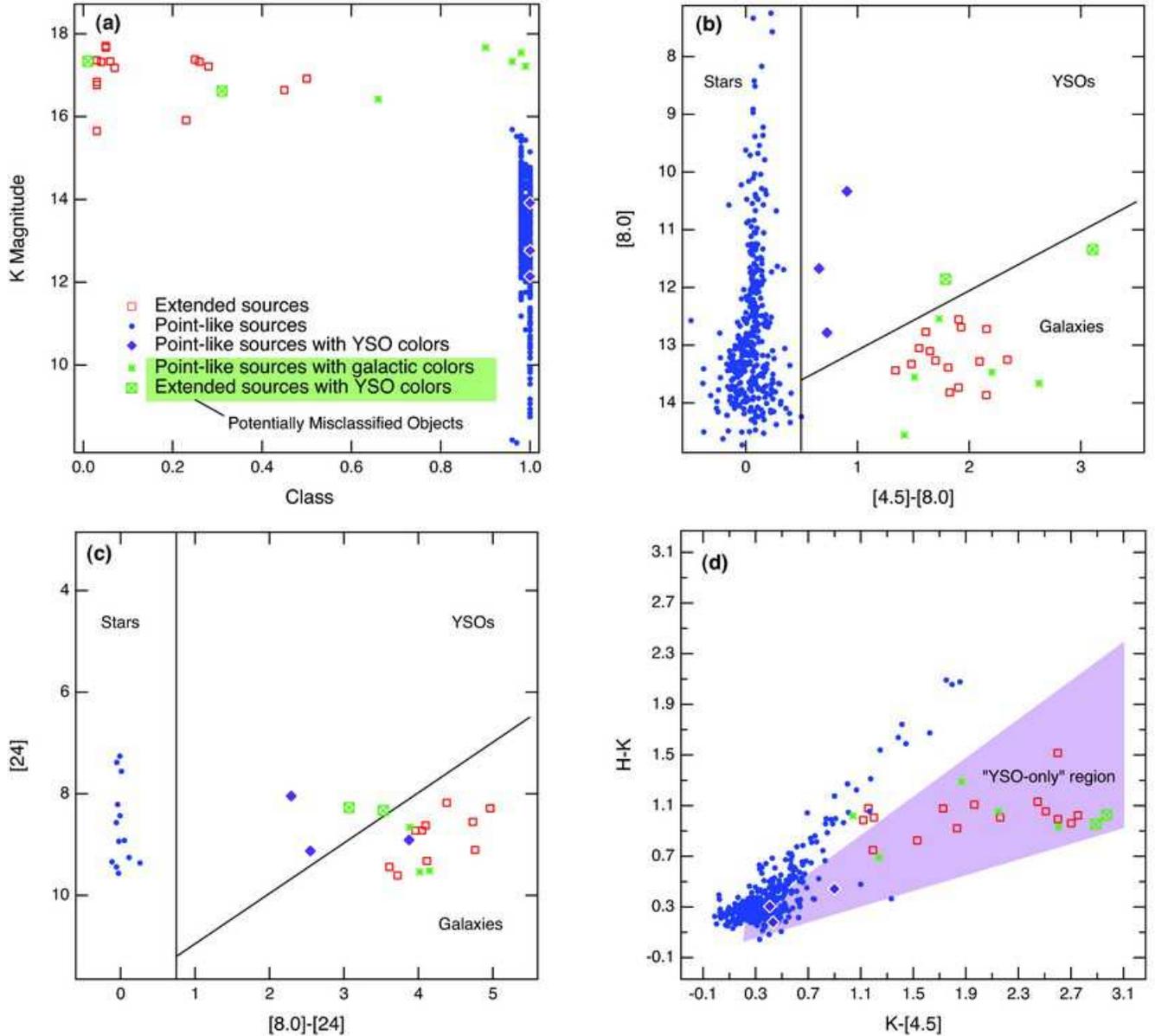}
\caption{Based on NIR data for six on-cloud frames we show (a) structural information for objects with high-quality detections in all four bands in IRAC which appear point-like; we resolve many as extended. Of the five objects identified as YSO candidates by \citet{Jorgensen:2006} based on either color-magnitude diagram (b) or (c), three are point-like and likely YSOs, and two are extended and probably galaxies. A handful of potentially point-like objects are either unresolved galaxies or low-luminosity YSOs previously unidentified. The addition of $H$ and $K$ photometry (d) is insufficient to distinguish between YSOs and galaxies. The shaded region is the area where ``only YSOs" are found in \citet{Jorgensen:2006} using 2MASS data, but this is because 2MASS only picks up bright local galaxies. With deeper near-infrared data, the ``YSO-only" area is littered with galaxies. }
\label{c2d}
\end{figure*}

No Spitzer data exist for our control fields, but  \citet{Jorgensen:2006} present c2d program data for all six on-cloud frames. Following their procedure for identifying YSOs, we cross-correlated our NIR data with the c2d catalog (DR3) and pulled out magnitudes for objects which were point-like with photometric quality ÔAÕ or ÔBÕ in all four IRAC bands. If a high-quality detection was made at 24 \micron (MIPS), that information was also included. Based on a comparison with a SWIRE extragalactic field, \citet{Jorgensen:2006} give two different color-magnitude cuts to identify YSOs. Points lying above and to the right of these cuts in either the [8.0] vs [4.5]-[8.0] or the [24] vs [8.0]-[24] color-magnitude diagram are considered candidate YSOs. These cuts are shown in Figure ~\ref{c2d}, along with our classification of objects as extended (open symbols) or point-like (filled symbols). The well-known Class 0 protostars in L1448 and B5 are excluded from this catalog as they have outflows making them appear extended in the IRAC images.

IRAC images are limited to a pixel scale of 1.2\arcsec\, and roughly 2\arcsec\ resolution. With our 0.45\arcsec\ pixel scale and 0.6\arcsec\ seeing, we are able to better identify marginally extended objects than Spitzer, as every object in Figure~\ref{c2d} is unresolved by Spitzer. The structural and color-magnitude divisions of objects into stellar and extended (galactic) objects matches exceedingly well for most of the 402 objects, but there are ten interesting objects. The \citet{Jorgensen:2006} criteria identify 5 candidate YSOs. Of these, three are point-like and bright; these are most likely genuine YSOs (purple diamonds in Figure~\ref{c2d}). Two candidate YSOs are significantly extended and thus more likely to be galaxies (green crossed-boxes in Figure~\ref{c2d}). Five objects unresolved even in the NIR have galaxy-like colors and magnitudes (green asterisks in Figure~\ref{c2d}). Of these, one is relatively extended, and the other four are all much fainter than the other point-like sources. We cannot definitively establish the nature of these objects---they could either be unresolved galaxies or low-luminosity YSOs. 

\citet{Jorgensen:2006} combined 2MASS with Spitzer data in both their Perseus data and an extragalactic survey and based on this suggest that a $H-K$ vs. $K-[4.5]$ color-color diagram should provide a clean separation between YSOs and galaxies. Unfortunately, this appears to be due to the magnitude limit of 2MASS, which only picks up local galaxies with star-like colors. With deeper NIR data, the region of the color-color diagram where Spitzer and 2MASS see only YSOs is actually heavily contaminated with galaxies (Figure~\ref{c2d}d). Structural, rather than photometric, information appears to be be required for making the YSO/galaxy distinction for sources fainter than $K$ = 15.

\section{Implications for Extinction Mapping:GNICER}

In studies of young clusters near-infrared excess sources constitute both signal (YSOs) and noise (everything else), but in extinction mapping using background stars all NIRX are potential contaminants. Brown dwarfs are, by and large, foreground sources, with colors unrelated to the cloud material. YSOs may be associated with ongoing star formation and appear in front of, within, or behind the cloud tracing none, some, or all of the cloud material. Other near-infrared excess sources are reddened by the cloud, but since their intrinsic color is redder than normal stars they will invariably produce an overestimate of the column density. 

However, as we have shown above, in regions relatively far from the galactic plane the majority of near infrared excess sources are galaxies, and therefore they are behind the cloud and potentially useful tracers of column density. As both the noise and resolution of extinction maps depend on the number of background sources, finding a way to use these galaxies is desirable. Indeed, in one way galaxies are an ideal probe, since they are never foreground to our cloud. \citet{Yasuda:2007} has recently used galaxy number counts from the Sloan Digital Sky Survey \citep{York:2000} as a probe of extinction. We seek to make use of galaxy NIR colors, which we expect to be a more sensitive measure.

\begin{figure}
\includegraphics*[scale=0.57]{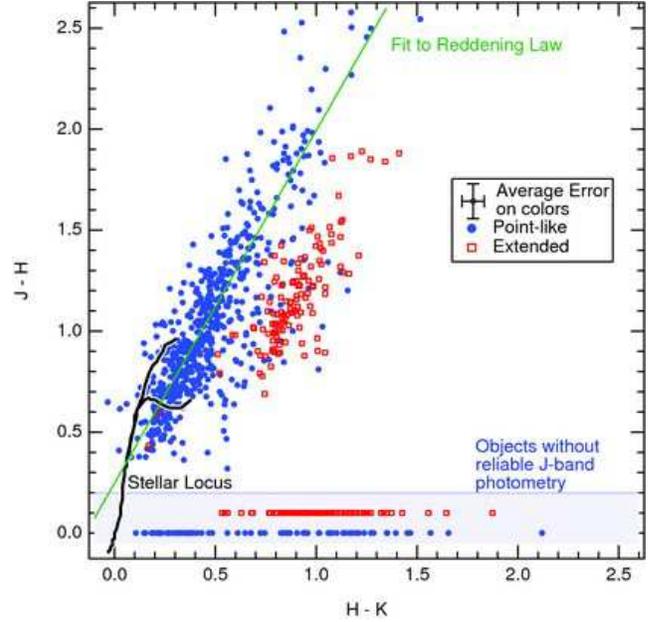}
\caption{L1451, a cloud with no known embedded young stars in the southwest of Perseus. Red open square are extended and blue closed symbols are point-like. Point-like and extended objects without a high quality J detection are plotted at J-H of 0 and 0.1 respectively to show their distribution in H-K. A fit to the reddening vector (shown) is consistent with that of \citet{Rieke:1985}.}
\label{L1451}
\end{figure}

We consider and compare four approaches to making an extinction map of L1451, a starless cloud in the southwest of Perseus: i) eliminating galaxies, ii) the naive treatment of blindly applying near-infrared extinction mapping to all sources, and iii-iv) two different methods for using galaxies. These different algorithms are summarized in Table~\ref{Algorithms}. The color-color diagram for L1451 is shown in Figure \ref{L1451} and in all cases we use Control 04 to determine the intrinsic properties. A $JHK$ color image of L1451 is shown in \citet{Foster:2006}.

\begin{deluxetable}{l|p{1.8 in}|l}
\tablecolumns{3}
\small
\tablewidth{0pt}
\tablecaption{Comparison of Different Extinction Algorithms}
\tablehead{\colhead{No.} & \colhead {Algorithm } & \colhead{Shown in Figures...}} 
\startdata
i & Discarding galaxies (NICER) & 14,15,16 (x-axis) 17 \\
ii & Treating all sources as stars & 14 (y-axis) 17 \\
iii & Using different mean colors and dispersions for galaxies and stars & 15a (y-axis)\\
iv & Same as (iii), but using magnitude-dependent properties for galaxies (GNICER) & 15b,16 (y-axis) 17\\
\enddata
\label{Algorithms}
\end{deluxetable}

\label{GNICERSection}

As shown in Figure \ref{L1451} it is not always easy to identify galaxies in a reddened field based simply on color. If we had $J$ band detections for every object, we could make a cut in color-color space parallel to the reddening vector and thus identify NIRX sources. However, reddening hopelessly mingles stars and galaxies if one only has their $H-K$ colors, as is often the case behind high column density regions. It is also possible to identify NIRX sources by using a preliminary extinction map to construct an intrinsic color-color diagram of all sources. This was the approach adopted by \citet{Lombardi:2006}. They constructed an initial extinction map for all their sources, converted observed colors into intrinsic, and eliminated near infrared excess sources. In our case, we are able to make a distinction based on structural information, with the ``sigma clipping" algorithm described below removing the few objects which are extended with normal colors, or point-like with near infrared excess.

Our baseline for comparison is an extinction map made with the standard application of the Near-Infrared Color Excess Method Revisited (NICER) after eliminating galaxies (Method i). Following the notation of \citet{Lombardi:2001}, each object provides a pencil beam measure of the extinction from
\begin{align}
A_V &= b_{J-H} [(J-H)^\mathrm{obs} - \mean{J-H}^\mathrm{con}] \label{estimator} \\ 
&+ b_{H-K}[(H-K)^\mathrm{obs} - \mean{H-K}^\mathrm{con}]. \notag
\end{align}
Assuming a normal reddening law, $b_{J-H}$ and $b_{H-K}$ are constrained such that 
\begin{equation}
\label{constraint}
b_{J-H} f_{J-H} + b_{H-K} f_{H-K} = 1
\end{equation}
where $f_{J-H}$ is the ratio of color excess to visual extinction, $f_{J-H} = E(J-H)/A_V = 1/9.35$ and likewise $f_{H-K} = 1/15.87$.

As the colors $J-H$ and $H-K$ are not independent (i.e. there is a stellar locus in the color-color diagram) and the photometric errors on the colors are also correlated (though we assume that errors on individual filters are uncorrelated) we calculate the covariance matrix $C = E + G$,
with
\begin{equation}
E = \left(
	\begin{array} { c c }
	\sigma_J^2 + \sigma_H^2 & -\sigma_H^2 \\
	-\sigma_H^2 & \sigma_H^2 + \sigma_K^2
	\end{array} \right),
\end{equation}
the covariance of the photometric errors and
\begin{equation}
G = \left(
	\begin{array} { c c }
	\Var(J-H) & \Cov(J-H,H-K)\\
	 \Cov(J-H,H-K) & \Var(H-K)
	\end{array} \right),
	\label{G}
\end{equation}
the intrinsic relationship in object colors measured from a control field. We wish to minimize the variance of Eqn.~\ref{estimator} subject to the constraint of Eqn.~\ref{constraint}. This is equivalent to (see \citet{Lombardi:2001} for full details) solving
\begin{equation}
\mathbf{b} = \frac{C^{-1} \cdot \mathbf{f}}{\mathbf{f} \cdot C^{-1} \cdot \mathbf{f}}.
\end{equation}
with $\mathbf{b} = (b_{J-H},b_{H-K})$ and $\mathbf{f} = (f_{J-H},f_{H-K})$. The error associated with each point like determination is
\begin{equation}
\Var(A_V) = \mathbf{b} \cdot C \cdot \mathbf{b}.
\label{pointerror}
\end{equation}

A map is created by smoothing these point estimates with a gaussian beam, weighting the contribution to a cell's extinction determination by the error on the object. Thus the error on a cell's extinction value is roughly inversely proportional to the square root of the number of point estimates used in that cell's determination. In the densest portions of the cloud, a significant number of background objects are not observable, and so contamination by foreground stars becomes a problem (typically this becomes a problem around A$_V$ = 10 mag \citep{Lombardi:2005}). To reduce this bias we perform ``sigma clipping". To do this, we determine a pixel's extinction, clip points more than three sigma deviant from the estimated A$_V$, and recalculate A$_V$ until this process converges. Where fewer than 5 objects contribute to a pixel's extinction determination we assign a blank value in the map and exclude that pixel from further analysis.

\begin{figure}
\includegraphics*[scale=0.55]{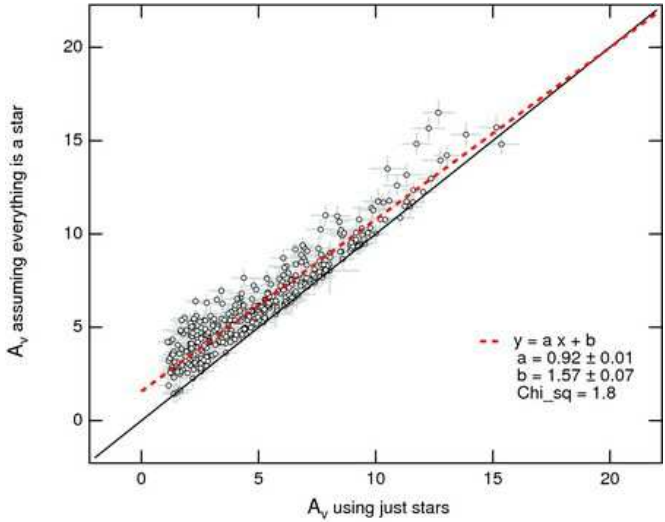}
\caption{A point-to-point comparison of an extinction map for L1451 derived naively assuming that every object has the mean color of the control field (Method ii), versus just using stars (i.e. point-like) objects (Method i). The solid line shows a one-to-one relation, and the dashed line shows a fit to the relation.}
\label{NaiveVstars}
\end{figure}

By rejecting extended objects in both our data frame and our control field property determination, we create our standard map for comparison (Method i). Figure \ref{NaiveVstars} shows the comparison between this map and the naive approach in which NIRX sources are not accounted for (Method ii), and so all sources are considered to be stars and de-reddened to the stellar locus. This naive approach is biased towards determining higher extinctions, as intrinsically red galaxies give large estimates of the column density. The problem is mitigated at high extinction because the galaxy population is intrinsically fainter than the stellar one and thus becomes a decreasing percentage of the sources seen through a thick cloud. Both these maps have a pixel size of 45\arcsec \ and no holes (i.e. pixels with less than 5 objects contributing).

Since we can separate stars and galaxies without recourse to their colors, galaxies provide crucial extra information in making an extinction map. We therefore introduce the Galaxies Near Infrared Color Excess method Revisited: GNICER. The basic approach (Method iii) is to divide objects into point-like and extended classes as before in both data and control fields and determine parameters for these populations separately. That is, solve for $\mean{J-H}^\mathrm{con}$ and $\mean{H-K}^\mathrm{con}$ for use in Eqn.~\ref{estimator} for both point-like and extended objects, and also determine the intrinsic covariance matrix $G$ in Eqn.~\ref{G} for both populations separately. Every source now gives a point-beam estimate with the uncertainty on that estimate (Eqn.~\ref{pointerror}) reflecting both the photometric error and how tightly such objects are intrinsically clumped in color-color space. Because our galaxies are approximately as tightly clumped as our stars, each galaxy provides roughly as much information about extinction as a star (subject to photometric uncertainty).

\begin{figure*}
\includegraphics*[scale=0.55]{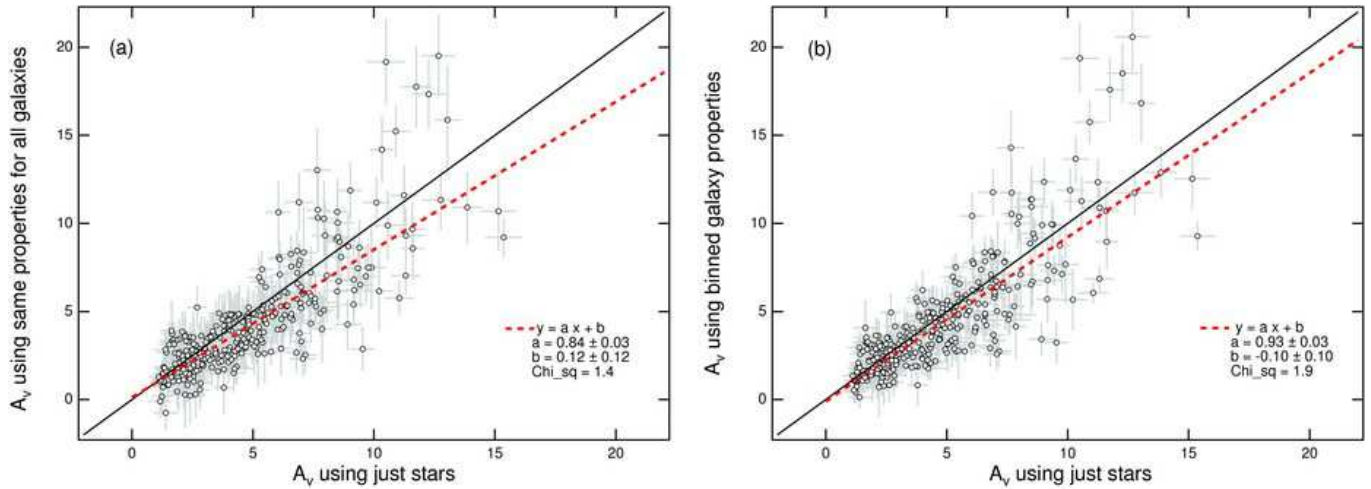}
\caption{A point-to-point comparison of an extinction map for L1451 as derived from \textit{just} galaxies versus just stars (Method i). Two different algorithms are compared. The simple treatment in (a) treats all galaxies as having the intrinsic properties of the control field galaxies (Method iii). The complex treatment in (b) assigns an intrinsic color to a galaxy based on its intrinsic magnitude using an extinction estimate from the stars (Method iv). This complex treatment helps remove bias at high A$_V$. The solid line shows a one-to-one relation, and the dashed line shows the fit.}
\label{TreatingGalaxies}
\end{figure*}

The most stringent requirement is that an extinction map created from only galaxies should provide roughly the same result as one created from just stars. The result of comparing GNICER (Method iii) for just galaxies with the straightforward application of NICER (for point-like sources only, Method i) is shown in Figure~\ref{TreatingGalaxies}a. Note that especially at larger A$_V$ the galactic estimate becomes quite scattered as there are fewer galaxies. As before, this analysis excludes pixels in the extinction map with fewer than five objects contributing to the estimate within that pixel (this is a number of pixels in the galaxy-map). The linear best-fit to the data (including errors in both X and Y) reveals a slope which is statistically significantly less than unity. Since the galaxies are detected only down to a brighter limit, this bias could arise from extinction structure within the pixel---the fainter galaxies trace low density material compared to the brighter stars in our magnitude limited image. Another source of this bias is the fact that brighter galaxies are preferentially bluer as fainter galaxies are typically further away, have a larger $k$-correction, and are redder. Behind significant extinction, we see only these bright, intrinsically bluer objects. Therefore, we estimate that they are sitting behind less extinction than they really are.

We test this second source of bias in our fourth experiment (Method iv) by splitting the control field galaxies into magnitude bins and determining the mean colors and color covariance matrix for galaxies up to a certain magnitude, $K_{lookup}$. We make a rough first extinction map using just stars. Then, in each pixel with extinction of A$_V$, galaxies are used to refine the extinction determination. For a galaxy of magnitude $K_{obs}$ we determine which set of galaxy properties from the control field to use by using

\begin{equation}
K_{lookup} = K_{obs} - 0.112 A_V.
\label{lookup}
\end{equation}

The properties of our galaxies binned this way is shown in Table \ref{GNICER}. Figure \ref{TreatingGalaxies}b show the result of this more complex algorithm, again comparing an extinction map made just with galaxies (y-axis, Method iv) against one made just with stars (x-axis, Method i). Most of the bias (non-unity slope) is removed by considering the color-magnitude relation of the galaxies. 

\begin{deluxetable}{cccccc}
\tablecolumns{6}
\tablecaption{GNICER parameters}
\tablehead{\colhead{Depth} & \colhead{\mean{J-H}} & \colhead{\mean{H-K}}
& \colhead{$G_{11}$} & \colhead{$ G_{12}$} & \colhead{$G_{22}$}
}
\startdata
   \sidehead{Stars}
   All  &  0.584 &    0.177  &    0.027  &    0.010  &    0.017  \\
  \sidehead{Galaxies}
   15.0   &  0.817 &   0.530   &    0.010  &    0.010  &    0.010 \\
   15.5   & 0.856  &   0.635   &   0.006   &   0.006   &   0.016 \\
   16.0   & 0.878  &   0.671   &   0.011   &   0.012   &   0.026  \\
   16.5   & 0.914  &   0.695   &   0.024   &   0.021   &   0.028 \\
   17.0   & 0.935  &   0.696   &   0.021   &   0.016   &   0.027 \\
   17.5   & 0.933  &   0.671   &   0.019   &   0.011   &   0.030 \\
\enddata
 \label{GNICER}
\tablecomments{Measured mean colors and color-covariance matrices of stars and galaxies binned by magnitude from Control 04. These parameters are used in GNICER to incorporate galaxies into an extinction map. $G_{11}=\Var(J-H), G_{12}=\Cov(J-H,H-K),G_{22}=\Var(H-K)$}
\end{deluxetable}

\begin{figure}
\includegraphics*[scale=0.55]{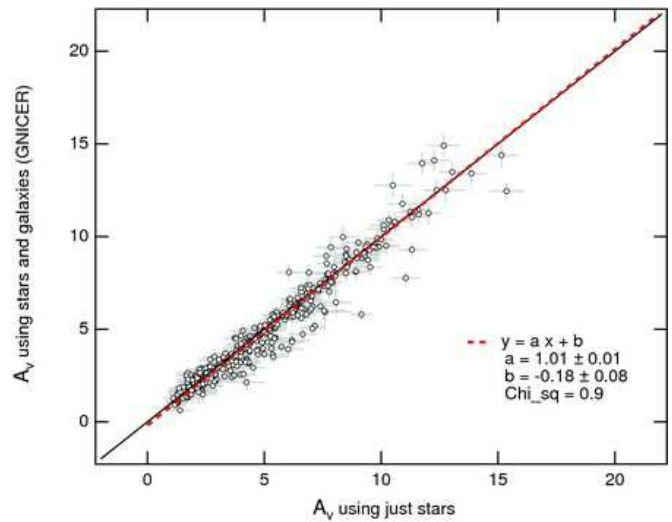}
\caption{A point-to-point comparison of an extinction map for L1451 derived by separating objects into two classes (extended and point-like) and using the magnitude-binned properties (mean color and dispersion) of these two populations as determined from the control field (Method iv), versus using just stars (Method i). In contrast to Figure~\ref{TreatingGalaxies}, the y-axis shows points from a map made from both stars and galaxies, rather than just galaxies. The solid line shows a one-to-one relation, and dashed line shows the fit.}
\label{CorrectVstars}
\end{figure}

\begin{figure}
\includegraphics*[scale=0.55]{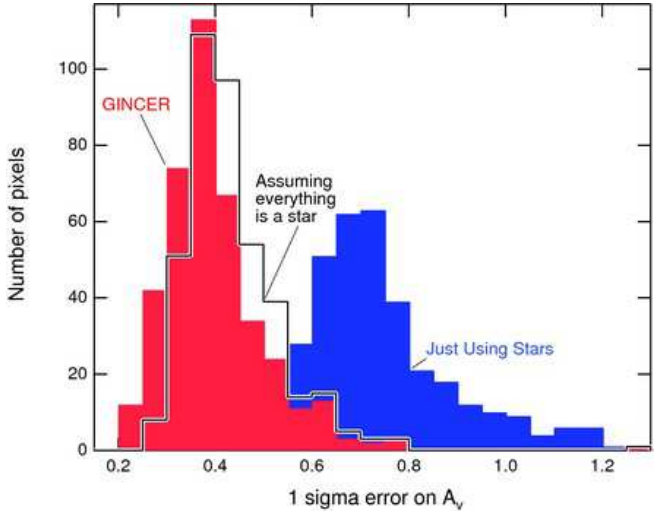}
\caption{A comparison of the map error amongst the three different way of making an extinction map (naively treating everything as a star, throwing out galaxies, and using galaxies). Errors on the naive determination (Method ii) are low, but this map is biased with a non-zero intercept and non-unity slope in comparison with a map where galaxies are discarded (Figure~\ref{NaiveVstars}, Method i). Including galaxies (GNICER, Method iv) gives many additional sources from which to estimate the extinction. This reduces the error without introducing significant bias (Figure \ref{CorrectVstars}). }
\label{ErrorComparison}
\end{figure}

When we combine the galaxy and stellar estimates in GNICER (Method iv) we make a map which is unbiased with respect to the map which eliminates galaxies (Method i, Figure \ref{CorrectVstars}). The big win is in the noise of the map, which is significantly reduced by adding galaxies as crucial extra estimators of the extinction. A comparison of the error distribution between naively assuming everything is a star (Method ii), rejecting galaxies (Method i), and this second algorithm for including galaxies in GNICER (Method iv) is shown in Figure \ref{ErrorComparison}. The naive treatment has relatively low estimated errors since it too uses all the background sources, but this is misleading because it is biased -- estimating much higher extinction for regions with low intrinsic extinction.

Future work which attempts to effectively use galaxies as additional sources for extinction mapping will ideally be able to: a) confidently categorize sources based on structural information (this requires adequate seeing, pixel scale, and image quality), and b) have control fields with adequate depth to determine the color-magnitude dependence of galaxies as seen through whatever portion of the galaxy is being studied. Galaxy clusters, and large-scale structure in the cosmological sense, pose a problem for GNICER. A galaxy cluster will be a dense group of objects at a small redshift range, perhaps significantly different from the average redshift for all galaxies. However, it is not clear that this problem is more severe than the problem which stellar clusters present NICER. Spatially coincident clumps of abnormally colored objects must be identified and understood in either case.

\section {Conclusion}

We have identified a number of near infrared excess sources (NIRX) in control fields in Perseus as normal galaxies at moderate redshifts. With high-quality images (i.e. good seeing conditions and fine pixel scale) we are able to classify objects as stars or galaxies based primarily on structural information. In such fields, galaxies which appear as NIRX sources constitute a significant background level in surveys for YSOs, and a significant potential contaminant in extinction maps using NICER. High-quality NIR imaging of cloud complexes observed in the mid-IR (e.g. with Spitzer) should provide the necessary discriminant to separate YSOs from galaxies in these data sets, extending the YSO luminosity function at the faint end where galaxy contamination currently dominates. This separation must be made from structural information, rather than relying on only colors. 

We have extended NICER to include galaxies in an unbiased way despite galaxies' significant color-magnitude relation (GNICER). Including galaxies in the analysis offers a significant increase in the number of background sources used to probe the structure of the cloud and a corresponding decrease in the uncertainty of this estimate. Using GNICER it should be possible to map low density clouds relatively far from the galactic plane with reasonable resolution. 

\acknowledgments

We thank Anthony Gonzalez for helpful comments, and Rosa Zapatero-Osorio and collaborators for allowing us to use their T-dwarf sequence ahead of publication. We thank the referee, Paul Harvey, for his helpful comments. This publication makes use of data products from the Two Micron All Sky Survey, which is a joint project of the University of Massachusetts and the Infrared Processing and Analysis Center/California Institute of Technology, funded by the National Aeronautics and Space Administration and the National Science Foundation. This material is based upon work supported by the National Science Foundation under Grant No. AST-0407172. EAL acknowledges support from NSF grant, AST02-02976, and NASA LTSA grant, NNG05D66G  to the University of Florida. JBF acknowledges support through NASA ADP grant NNG05GC39G.

{\it Facilities:} \facility{CAO:3.5m ()}

\bibliographystyle{apj}
\bibliography{NIRX}

\clearpage

\end{document}